# HETEROGENEOUS HIGHLY PARALLEL IMPLEMENTATION OF MATRIX EXPONENTIATION USING GPU

Chittampally Vasanth Raja, Srinivas Balasubramanian, Prakash S Raghavendra

Department of Information Technology, National Institute of Technology Karnataka, Surathkal, India

ch.vasanth.raja@gmail.com, abishek3000@gmail.com, srp@nitk.ac.in

## ABSTRACT

*The vision of super computer at every desk can be realized by powerful and highly parallel CPUs or GPUs or APUs. Graphics processors once specialized for the graphics applications only, are now used for the highly computational intensive general purpose applications. Very expensive GFLOPs and TFLOP performance has become very cheap with the GPGPUs.*

*Current work focuses mainly on the highly parallel implementation of Matrix Exponentiation. Matrix Exponentiation is widely used in many areas of scientific community ranging from highly critical flight, CAD simulations to financial, statistical applications. Proposed solution for Matrix Exponentiation uses OpenCL for exploiting the hyper parallelism offered by the many core GPGPUs. It employs many general GPU optimizations and architectural specific optimizations. This experimentation covers the optimizations targeted specific to the Scientific Graphics cards (Tesla–C2050). Heterogeneous Highly Parallel Matrix Exponentiation method has been tested for matrices of different sizes and with different powers. The devised Kernel has shown 1000X speedup and 44 fold speedup with the naive GPU Kernel.*

## KEYWORDS

*Matrix Exponentiation, GPGPU, OpenCL, Highly Parallel Matrix Exponentiation*

## 1. INTRODUCTION

Power consumption and speed of light limitation have become the limiting factors for the moorely's law. These factors have opened doors for the dawn of the multi core processors. All processors manufacturing companies are moving towards multi core processors development. The problem with the multi core processors is that despite we have multiple cores of some GHz processing frequency we if run a single thread applications we get only one core processing power. In order to increase the processing speed and thereby increasing the throughput of our applications, we have to write applications by keeping the multi core architecture in mind.

Heterogeneous computing and specialized computing units have started a new age of high speed computation. The processing power of every workstation continues to grow at an exponential rate. Exploitation of the processing power made available by the parallel computing revolution is essential to enhance the user experience.

Specialized processors like the GPU have now made their computational power available to their computational power available to the common man. GPGPU has drastically changed the landscape of heterogeneous computing. OpenCL is a much favoured language for heterogeneous computing and is soon emerging as an industrial standard. Matrix exponentiation has wide variety of applications in the scientific community.

We look into the task of parallelizing this task for efficient execution on the Tesla 2050C graphics card using OpenCL. We use a lesser number of kernel execution to boost up the performance. This method not only increases the throughput significantly as the power of the matrix increases.

## 2. RELATED WORK

Many matrix operations have been the basic operations in all of walks the scientific community. Matrix – matrix multiplication, Matrix power computations are very pivotal and require more amount of time as the size of the matrix increases. In order to boost up the performance of these time consuming operations many have proposed GPU based implementations. CUDA BLAS library is one such basic implementation. The problem with the CUDA BLAS library is that CUDA works only with the NVIDIA graphic cards and it will not work with other computing devices. Many optimizations to the Matrix operations have been proposed. [6] Volkov and Demmel present an experimental study of GPU memory subsystem and an efficient implementation of dense matrix-matrix multiplication. The implementation is shown to be nearly optimal under the constraints of hardware implementations. Yinghong Sun and Yuanman Tong [7] proposed several optimizations to the highly fine grained matrix multiplication and matrix vector multiplications. This implementation is again based on CUDA, hence is not heterogeneous. This implementation also lacks the some other improvements like TILING and architecture specific optimization. Our method exploits all the optimizations and architecture specific optimizations. Our solution was tested thoroughly for the precision problem and high matrix sizes up to 512 by 512.

The rest of the paper is organized in the following way. Section 3 presents GPU architecture and OpenCL programming model. Section 4 presents our methodology and optimizations incorporated in detail. In Section 5, Experimental setup and results were presented. Finally, conclusion is presented in Section 6.

## 3. GPU ARCHITECTURE AND OPENCL PROGRAMMING MODEL

### 3.1 GPU Architecture:

As in Central Processing Units (CPUs), Graphics processing Units (GPUs) also come with wide verities of architectures and options. Here we discuss a formal and widely accepted fundamental architecture.

GPU parallel computing architecture comprises massively multi threaded capable sophisticated hardware and also sophisticated hardware task management because early GPUs are designed to processes high end graphic workloads. GPUs comprise of set of parallel multi processors which are also called as Streaming multi processors. Each Streaming multi processor intern contains a set of processor cores as shown in Figure 1.

Thread scheduler shown in the Figure 1 is a hardware based thread scheduler. This thread scheduler is responsible for the thread scheduling across the thread processing clusters. This achieves nearly 100 % utilization. If a thread is waiting for a memory access, a scheduler can perform the zero cost immediate context switches to another thread. Since the thread creation management are taken care by a dedicated hardware not only many threads can be created also can be easily maintained and scheduled efficiently.

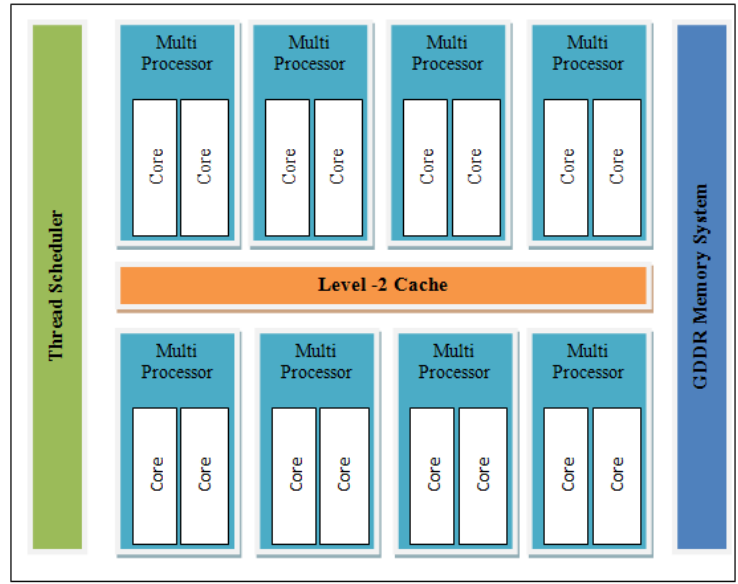

Figure 1. GPU Architecture

### 3.1.1 GPU Memory Model:

GPUs also contain different levels of memory model. The hierarchy goes like this, 1) Private memory/Registers, 2) Local memory or On-chip scratch pad memory, 3) Shared Global /constant memory data cache, 4) Global memory, 5) Host memory. These memories are shown in the Figure 3.

All the above mentioned memories operate at different clock rates; usually the memories which are much closer to the processor or core are faster similar to CPUs. In the above mentioned memory order, as we move from 1) Registers to 5) Host memory, time cycles for accessing data in that memory increases hence memory latency increases. GPU Registers are similar to CPU registers and operate with very few clock cycles usually 1-2 clock cycles. Local memory or On-chip scratchpad memory can be used by all the work items with in the work group. This local memory is also a fast accessible memory. The scratch pad memory is software managed unlike the cache. So the scratch pad memory can be optimally programmed by the developer to get the best performance. Shared Global memory data cache, this a hardware managed cache for global and constant memory. The constant memory is also a part of global memory. Global memory contains a very large memory size with very high latency. Global memory can be used my all the compute units in the Device.

### 3.2 OpenCL programming model:

Open Computing Library (OpenCL) is an open source, royalty free, heterogeneous computing parallel programming language. Heterogeneity means applications written in OpenCL can run on any hardware including CPUs, GPUs, APUs or any other accelerators. OpenCL allows both task and data parallelisms.

OpenCL platform model is shown in the Figure 2. The Host can be Personal computer, Super computer or even embedded system, which provides OpenCL API and run time compiler. Compute device can be CPU, GPU, APU, DSP, Cell Processor or any other accelerator which can execute the OpenCL kernel written in C99 programming language.

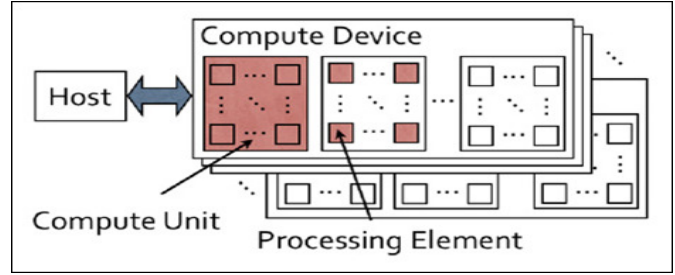

Figure 2. OpenCL Platfrom model

OpenCL's APIs include functionality for 1) Finding available target devices (CPUs, GPUs, and various accelerators) 2) managing the target devices' contexts; 3) Managing memory allocations; 4) Performing host-device memory transfers; 5) Compiling the OpenCL programs and kernel functions that the devices will execute; 6) launching kernels on the target devices; 7) querying execution progress; and 8) checking for errors. The above functionality allows the OpenCL programming model to treat the host device and accelerated device to be programmed, as heterogeneous. Figure 4 summarizes OpenCL programming steps with the above specified OpenCL API functionality.

### 3.2.1 OpenCL Memory Model:

All the data transfers between the Host and accelerating device are done explicitly by the developer.

OpenCL Supports On-chip scratchpad memory as discussed in the previous GPU architecture. Since the scratch pad memory is software managed, developer has to fully manage this memory explicitly.

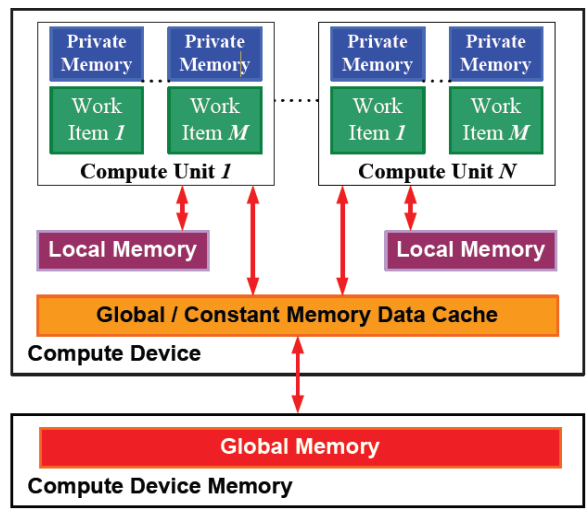

Figure 3. OpenCL Memory model

OpenCL allows developer to write code to run on a compute device in a special way called as kernel code. The kernel is written in OpenCL C99 language. The kernel code can be compiled, linked and run on compute device on the fly or it can allow it to have offline compiled code to be run on the compute device. The former way of run time compilation is recommended, since features specific to the under lying hardware can be easily employed. Runtime compilation

108

eliminates dependencies on the hardware and its instruction set hence hardware vendors can easily change their instruction set, drivers and supporting libraries.

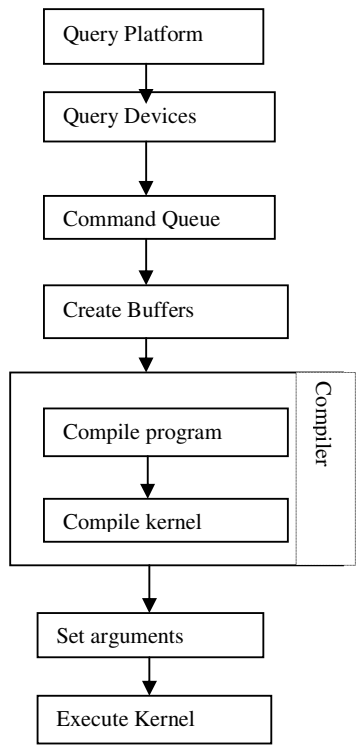

Figure 4: OpenCL Programming Steps

## 4. IMPLEMENTATION DETAILS

In this section we compare the naive way of matrix multiplication with our methodology.

### 4.1 Naive CPU method:

**Algorithm:** Matrix Exponentiation

**Module:** Matrix Multiplication (Matrix, Size)

```
for i = 1 to n do
 do
   for j = 1 to n do
    do
      for k = 1 to n do
       do
              c[i,j] = c[i,j] + a[i,k] * b[k,j];
       done
    done
 done
```

Call the above function to the power times. This step essentially adds one more loop.

### 4.2 Naïve GPU mechanism:

Call the GPU kernel N times from the host code to multiply the given matrix N times.

## 4.3 Our Approach:

Our methodology doesn't multiply the matrix N (given power) times. Instead it multiplies the given matrix log(N) times. This reduces the number of multiplications drastically. Our method uses the highly optimized GPU kernel for the matrix multiplication. Since the matrix multiplication is highly data parallel application it gives the maximum speed up in a fine grained parallelism supported devices like GPUs. Matrix multiplication is a well suited algorithm for SIMD nature stream processors.

In addition to mathematically optimized kernel matrix multiplication it also uses optimizations specific to the GPUs.

- Effective usage of Work Groups
- Cache optimization and local memory
- Coalesced memory reads / writes
- Loop unrolling
- Architecture aware register utilization
- Effective utilization of barriers
- Usage of TILING concept
- Less amount of data transfer between Host and GPU

### 4.3.1 Effective usage of Work Groups:

OpenCL enables us to group the work items with in the same work group. All the work items with in the same work group can share the data. This decreases amount of global data transfer and increases the performance. Within the same work group the synchronizations is not implicitly done it has to be done by the developer. Among different work groups synchronizations is implicitly done. This solution uses the (ROW / 4 * COL / 4) global Work items and 32 by 32 local work items with in each work group.

### 4.3.2 Cache optimizations and local memory:

In our approach to matrix exponentiation we use the less capacity local memory 16KB very efficiently. Usage of local memory increases the performance because each local memory access takes a very less magnitude of clock cycles in the range of (2 – 10).

### 4.3.3 Coalesced global memory reads / writes:

Our method has been written to exploit the Coalesced memory optimized reads and writes. Since the matrix elements are stored in the Row Major Order, data required for each thread are implemented in a coalesced reads fashion and data written after computations are of coalesced writes. This reduces the high cost of reading from and writing to the global memory.

### 4.3.4 Loop Unrolling:

Loop unrolling is one another optimization technique employed with the factor of 4, 8, 16 with respect to the matrix size.

### 4.3.5 Architecture aware register utilization:

GPU is SIMD in nature hence usage of vector data types gives better performance. We used vectors of size 4. This has improved the performance up to 3%.

### 4.3.6 Effective utilization of barriers:

Barriers are used for synchronization of memories in local and global levels. If barriers are not properly used performance comes down some times our GPU computation performance may go down than the sequential code, and if not used at all, may lead to the inconsistence of data. In this project we used the local memory synchronization barriers to make sure the data is consistent with in each work group.

### 4.3.7 TILED Matrix Multiplication:

It uses TILED matrix multiplication method to decrease the high global data transfers and increase the utilization of local memory. This project comprises different kernel having different TILES of size 4 * 4, 4 * 8, 8 * 8, 16 * 8, 8 * 16 and 16 * 16. An appropriate TILE size is used based on the problem and local memory available. In our solution on Tesla C 2050 graphics card we evaluated using 16 * 16 TILE size.

### 4.3.8 Less amount of data transfer between Host and GPU:

The amount of data transfer between the Host and GPUs gets reduces drastically because the data is offloaded only log(N) times.

## 5. EXPERIMENTAL RESULTS

### 5.1 Experimental Setup:

For this experimental set up Intel Xeon process was used having 16 cores, each having 2.40 GHz clock frequencies. Host system having primary memory of 8 GB is used. The graphics card used is Tesla C2050 having 448 each having clock frequency of 1150 MHz. Detailed Technical description is given in Table 1. The code was compiled using OpenCL 1.1 NVIDIA CUDA version.

**TABLE 1. NVIDIA TESLA C2050 SPECIFICATIONS**

| Model of GPU | NVIDIA Tesla C 2050 |
|---|---|
| Number of Processors | 14 |
| Number of cores | 448 |
| Number of cores per Processor | 32 |
| Clock Frequency | 1150 (in MHz) |
| Core clock Frequency | 575 (in MHz) |
| Bandwidth | 144 (GBs/Sec) |
| Bus Type | GDDR5 |
| Processing Power max in GFLOPs | 1288 |

### 5.2 Results

Results obtained for the above methodology of number of matrix multiplication showed very higher performance improvement. In the following section presents comparison of Naive GPU matrix exponentiation method with Our methodology. We compare both the methods with matrices of various sizes and increasing powers.

Table 2. Exponentiation of Matrix of Size 64 by 64

|  | 64 | 128 | 256 | 512 | 1024 |
|---|---|---|---|---|---|
| **Naïve GPU** (In Sec) | 0.05 | 0.14 | 0.43 | 0.99 | 2.69 |
| **Sequential CPU** (In Sec) | 0.23 | 0.68 | 1.74 | 4.31 | 10.83 |
| **Naïve Speed UP** | **4.6** | **4.86** | **4.05** | **4.35** | **4.03** |
| **Our Approach** (In Sec) | 0.01 | 0.01 | 0.02 | 0.02 | 0.03 |
| **Our Approach vs Naïve GPU** | 5 | 13.99 | 21.48 | 49.54 | 89.58 |

The above results can be graphically represented in the following graphs.

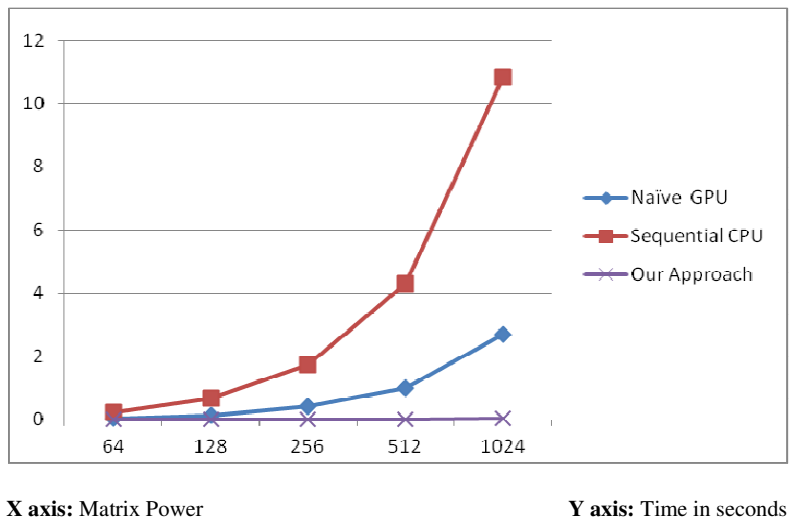

**X axis:** Matrix Power　　　　　　　　　　　**Y axis:** Time in seconds

Figure 5. Performance comparison of Naïve GPU Kernel, Sequential CPU and Our approach for the matrix of size 64 by 64

From Figure 5, it is evident that that Naive GPU is having very good performance speed up of almost 4 times. Even though the power of matrix is increasing exponentially the Naive method speedup remained constant. Our methodology of the matrix multiplication has not only showed the high performance improvement over the Naive GPU method but also speeded up accordingly with the exponential increase of matrix power.

Figure 6 shows the comparison of speedups with Naive kernel vs. Our approach for the matrix of Size of 64 by 64 with respect to Sequential CPU matrix exponentiation implementation. It is clear from the above bar chart that our methodology has not only improved from the Sequential CPU matrix multiplication but also improved significantly than the Naive GPU implementation.

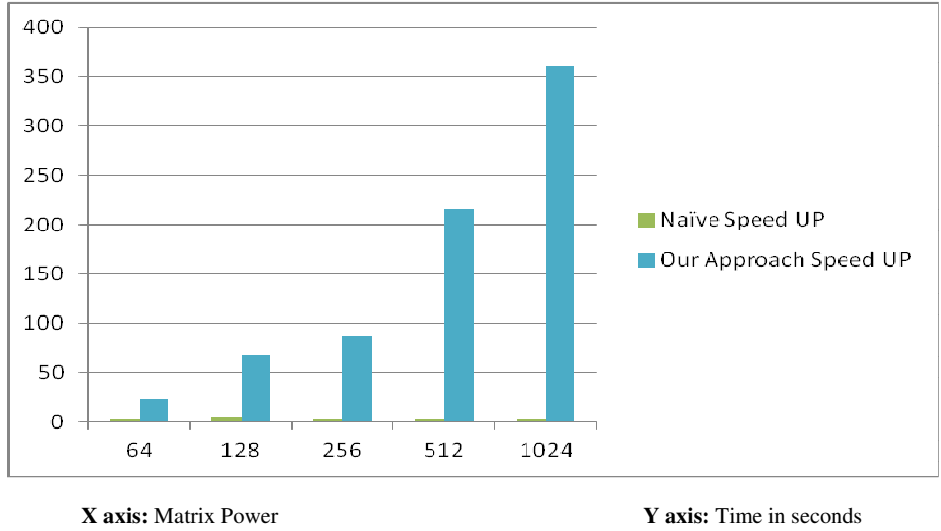

**X axis:** Matrix Power          **Y axis:** Time in seconds

Figure 6. Comparison of Speedup achieved Naïve kernel vs. our approach for the matrix of size 64 by 64

**Table 1 Exponentiation of Matrix of Size 128 by 128**

|  | 64 | 128 | 256 | 512 |
|---|---|---|---|---|
| **Naïve GPU** (In Sec) | 0.1 | 0.25 | 0.62 | 1.38 |
| **Sequential CPU** (In Sec) | 1.83 | 5.72 | 13.18 | 27.53 |
| **Naïve Speed UP** | 18.3 | 22.88 | 21.26 | 19.95 |
| **Our Approach** (In Sec) | 0.02 | 0.02 | 0.02 | 0.02 |
| **Our Approach vs. Naïve GPU** | 5 | 12.5 | 31 | 69 |

The above results can be graphically represented in the following chart.

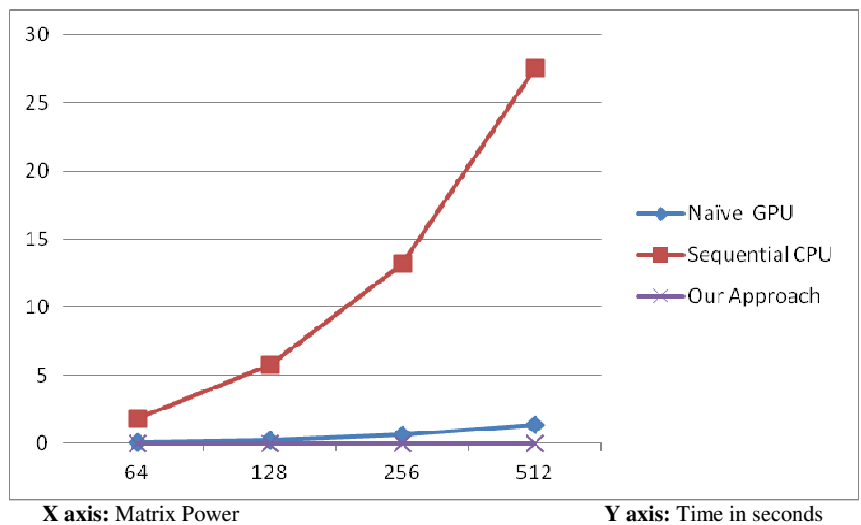

**X axis:** Matrix Power          **Y axis:** Time in seconds

Figure 7. Performance comparison of Naïve GPU Kernel, Sequential CPU and Our approach for the matrix of size 128 by 128

Figure 7, shows Naive GPU is having very good performance speed up of almost 18 fold for the matrix of size 128 by 128. Despite the power of matrix increasing exponentially the Naive method speedup remained constant i.e., 18. Our methodology of the matrix multiplication has not only showed the higher performance improvement over the Naive GPU method but also speeded up accordingly with the exponential increase of matrix power.

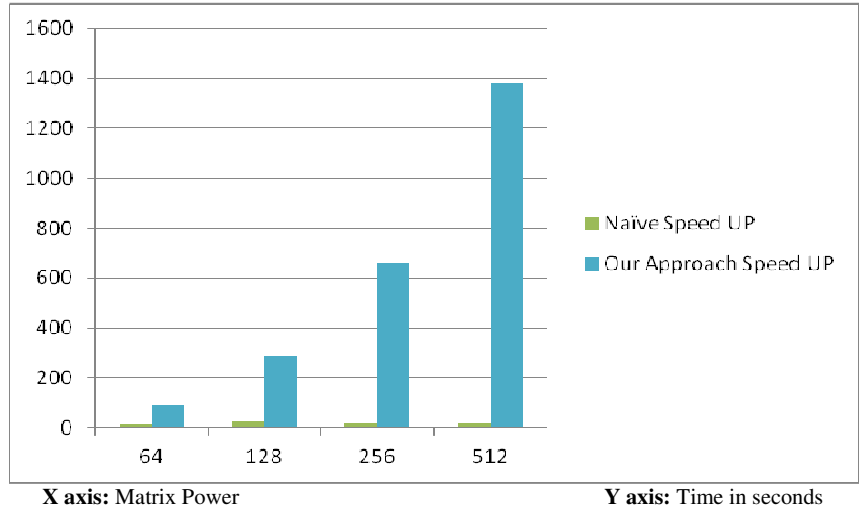

**X axis:** Matrix Power                    **Y axis:** Time in seconds

Figure 8. Comparison of Speed UP achieved Naïve kernel vs. our approach for the matrix of size 128 by 128

Figure 8 depicts the comparison of speedups with Naive kernel vs. our approach for the matrix of Size of 128 by 128 with respect to Sequential CPU matrix exponentiation implementation. It is clear from the above bar chart that our methodology has not only improved from the Sequential CPU matrix multiplication but also improved significantly than the Naive GPU implementation.

**Table 2 Exponentiation of Matrix of Size 256 by 256**

|  | 64 | 128 | 256 | 512 |
|---|---|---|---|---|
| **Naïve GPU** (In Sec) | 0.21 | 0.43 | 0.87 | 1.76 |
| **Sequential CPU** (In Sec) | 16 | 32.19 | 64.61 | 129.38 |
| **Naïve Speed UP** | 76.19 | 74.86 | 74.26 | 73.51 |
| **Our Approach** (In Sec) | 0.03 | 0.03 | 0.04 | 0.04 |
| **Our Approach vs. Naïve GPU** | 7 | 14.33 | 21.75 | 44 |

The above results can be graphically represented in the following graphs

International Journal of Distributed and Parallel Systems (IJDPS) Vol.3, No.2, March 2012

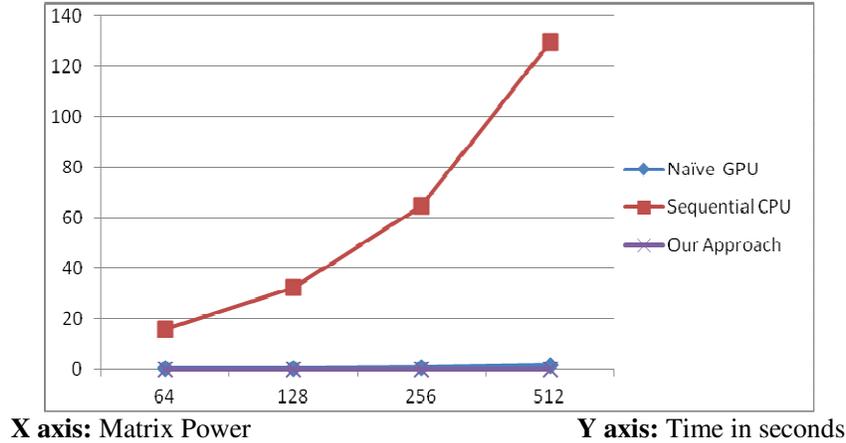

**X axis:** Matrix Power             **Y axis:** Time in seconds

Figure 9. Performance comparison of Naïve GPU Kernel, Sequential CPU and Our approach for the matrix of size 256 by 256

Figure 9 shows Naive GPU is having very good performance speed up of almost 73 fold for the matrix of higher size 256 by 256. Despite the power of matrix increasing exponentially the Naive method speedup remained constant i.e., 73. Our methodology of the matrix multiplication has not only showed the higher performance improvement over the Naive GPU method but also speeded up accordingly with the exponential increase of matrix power. The dense matrix of size 256 by 256 with the matrix high exponentiation i.e., power 512 has shown a tremendous power improvement of 44 times speedup for our method than the Naive GPU kernel code.

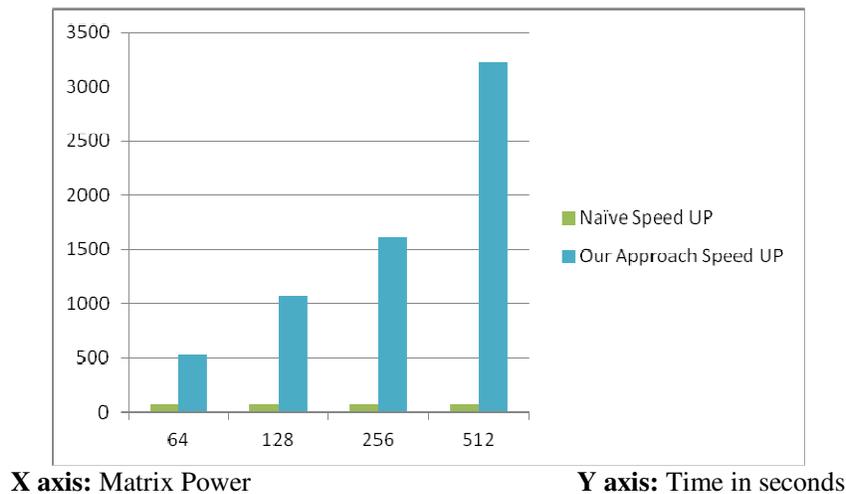

**X axis:** Matrix Power             **Y axis:** Time in seconds

Figure 10. Comparison of Speed UP achieved Naïve kernel vs. our approach for the matrix of size 256 by 256

Figure 10 depicts the comparison of speedups with Naive kernel vs. Our approach for the matrix of Size of 256 by 256 with respect to Sequential CPU matrix exponentiation implementation. It is clear from the above bar chart that Our methodology has not only improved from the

115

Sequential CPU matrix multiplication but also improved significantly than the Naive GPU implementation.

**Table 3 Exponentiation of Matrix of Size 512 by 512**

|  | 64 | 128 | 256 |
|---|---|---|---|
| **Naïve GPU** (In Sec) | 0.26 | 0.43 | 0.87 |
| **Sequential CPU** (In Sec) | 78.49 | 157.62 | 315.74 |
| **Naïve Speed UP** | 301.88 | 366.56 | 362.92 |
| **Our Approach** (In Sec) | 0.12 | 0.13 | 0.14 |
| **Our Approach vs. Naïve GPU** | 2.16 | 3.31 | 6.21 |

The above results can be graphically represented in the following graphs

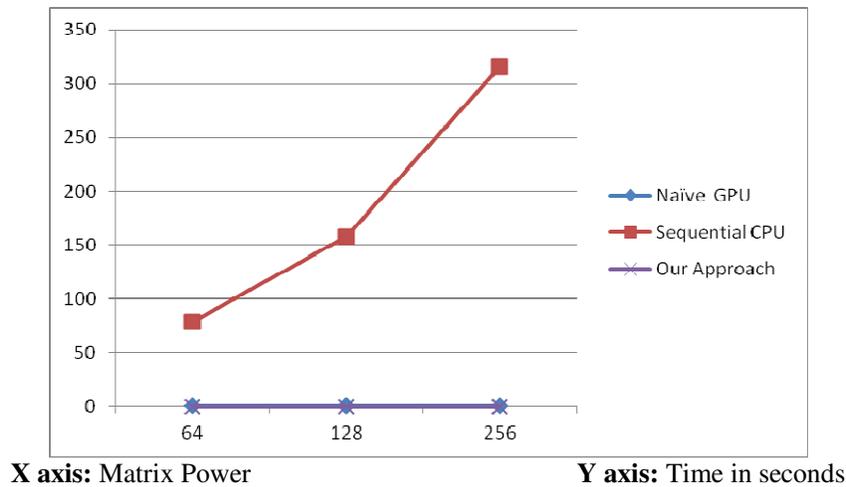

**X axis:** Matrix Power  **Y axis:** Time in seconds

Figure 11. Performance comparison of Naïve GPU Kernel, Sequential CPU and Our approach for the matrix of size 512 by 512

Figure 11 shows Naive GPU methodology having performance speed up of almost 300 fold for the matrix of bigger size 512 by 512. Our methodology of the matrix multiplication has not only showed the higher performance improvement over the Naive GPU method but also speeded up accordingly with the exponential increase of matrix power. The dense matrix of size 512 by 512 with the matrix high exponentiation i.e., power 256 has shown a tremendous performance improvement of 6 times speedup for our method than the Naive GPU kernel code.

Figure 12 depicts the comparison of speedups with Naive kernel vs. our approach for the matrix of Size of 512 by 512 with respect to Sequential CPU matrix exponentiation implementation. It is clear from the above bar chart that our methodology has not only improved from the Sequential CPU matrix multiplication but also improved significantly than the Naive GPU implementation.

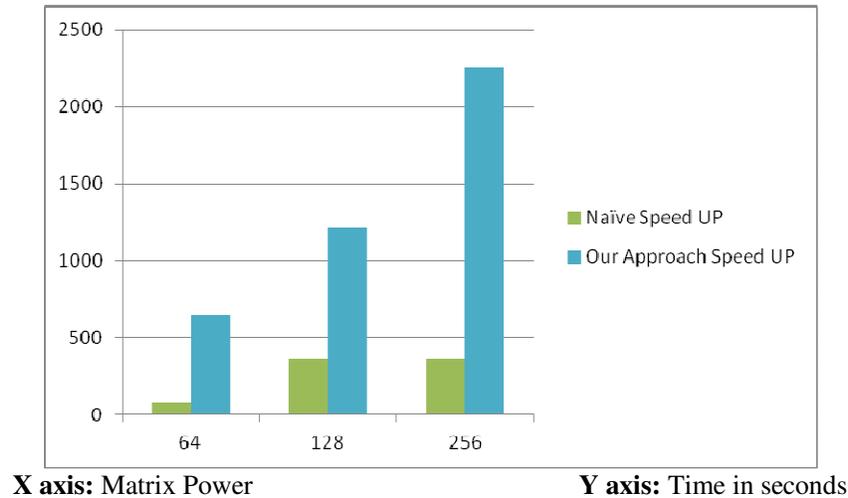

**X axis:** Matrix Power                    **Y axis:** Time in seconds

Figure 12. Comparison of Speed UP achieved Naïve kernel vs. our approach for the matrix of size 512 by 512

## 6. CONCLUSION

Current methodology is implemented in heterogeneous language OpenCL we can run on any compute device on any architecture. In this experiment we have tested our algorithm on dense matrices of size up to 512 by 512 against higher powers up to 256 and evaluated the results. All the results are strictly compared with the sequential code results for any precision problems. Our methodology preserves the high precision and enables the supercomputing capability with the relatively cheaper GPUs.

Our solution gives more than thousand fold performance on the high end scientific graphic card Tesla C 2050 for the higher power of matrices of bigger sizes. This approach includes several architectural performance benefits specific to Tesla C 2050 and also some general optimization techniques supported by all multi core processors including GPUs. In Fig 5 to Fig 12, our approach is compared with the Naïve GPU method and our method always outperforms the Naïve GPU approach.

International Journal of Distributed and Parallel Systems (IJDPS) Vol.3, No.2, March 2012

**Authors**

**Name:** Chittampally Vasanth Raja.

Currently pursuing M.Tech in the Department of Information Technology, National Institute of Technology Karnataka, Surathkal. His areas of interests include High performance computing, SOA, Image processing.

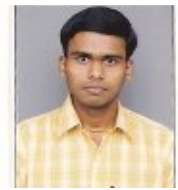

International Journal of Distributed and Parallel Systems (IJDPS) Vol.3, No.2, March 2012

**Name:** Srinivas Balasubramanian.

Currently pursuing M.Tech Research in the Department of Information Technology, National Institute of Technology Karnataka, Surathkal. His areas of interest includes High performance computing, Design and Analysis of Algorithms.

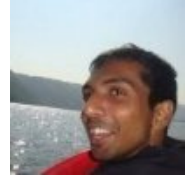

**Name:** Dr. Prakash S Raghavendra.

Currently working as an Asst Professor in the Department of Information Technology, National Institute of Technology Karnataka, Surathkal. His areas of interest include  High Performance Computing, Compiling for Distributed Memory Machines,  Compiling for SPMD programs, Performance Analyses Techniques (Profiling Tools), Parallel and Distributed Computing, High Performing Virtual Machines, GPGPU Computing and its applications, Rich Internet Applications (RIA), Performance of RIAs, Automatic generation of Rich internet clients.

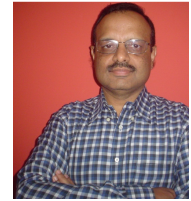